\documentclass[prl,superscriptaddress,floatfix,twocolumn]{revtex4}
\usepackage{amsfonts}
\usepackage{stmaryrd}
\usepackage{bbm}
\usepackage{mathrsfs}
\usepackage{tipa}
\usepackage{amssymb}
\usepackage{txfonts}
\usepackage{graphicx}
\usepackage{dcolumn}
\usepackage{epstopdf}
\usepackage{array}
\usepackage{color}

\newcommand{\PreserveBackslash}[1]{\let\temp=\\#1\let\\=\temp}
\newcolumntype{C}[1]{>{\PreserveBackslash\centering}p{#1}}
\newcolumntype{R}[1]{>{\PreserveBackslash\raggedleft}p{#1}}
\newcolumntype{L}[1]{>{\PreserveBackslash\raggedright}p{#1}}

\begin{document}

\newcommand*{\cm}{cm$^{-1}$\,}

\title{Signature of weakly coupled $f$ electrons and conduction electrons in magnetic Weyl semimetal candidates PrAlSi and SmAlSi}

\author{Rui Lou}
\thanks{lourui@lzu.edu.cn}
\affiliation{School of Physical Science and Technology, Lanzhou University, Lanzhou 730000, China}
\affiliation{Leibniz Institute for Solid State and Materials Research, IFW Dresden, 01069 Dresden, Germany}
\affiliation{Helmholtz-Zentrum Berlin f{\"u}r Materialien und Energie, Albert-Einstein-Stra{\ss}e 15, 12489 Berlin, Germany}
\affiliation{Joint Laboratory ``Functional Quantum Materials" at BESSY II, 12489 Berlin, Germany}

\author{Alexander Fedorov}
\thanks{a.fedorov@ifw-dresden.de}
\affiliation{Leibniz Institute for Solid State and Materials Research, IFW Dresden, 01069 Dresden, Germany}
\affiliation{Helmholtz-Zentrum Berlin f{\"u}r Materialien und Energie, Albert-Einstein-Stra{\ss}e 15, 12489 Berlin, Germany}
\affiliation{Joint Laboratory ``Functional Quantum Materials" at BESSY II, 12489 Berlin, Germany}

\author{Lingxiao Zhao}
\thanks{zhaolx@mail.sustech.edu.cn}
\affiliation{Department of Physics, Southern University of Science and Technology, Shenzhen 518055, China}

\author{Alexander Yaresko}
\affiliation{Max Planck Institute for Solid State Research, 70569 Stuttgart, Germany}

\author{Bernd B{\"u}chner}
\affiliation{Leibniz Institute for Solid State and Materials Research, IFW Dresden, 01069 Dresden, Germany}
\affiliation{Institute for Solid State and Materials Physics, TU Dresden, 01062 Dresden, Germany}

\author{Sergey Borisenko}
\thanks{s.borisenko@ifw-dresden.de}
\affiliation{Leibniz Institute for Solid State and Materials Research, IFW Dresden, 01069 Dresden, Germany}

\begin{abstract}
  Magnetic topological materials are a class of compounds with the underlying interplay of nontrivial band topology and magnetic spin configuration. Extensive interests have been aroused due to their application potential involved with an array of exotic quantum states. With angle-resolved photoemission spectroscopy and first-principles calculations, here we study the electronic properties of two magnetic Weyl semimetal candidates PrAlSi and SmAlSi. Though the two compounds harbor distinct magnetic ground states (ferromagnetic and antiferromagnetic for PrAlSi and SmAlSi, respectively) and 4$f$ shell fillings, we find that they share quite analogous low-energy band structure. By the measurements across the magnetic transitions, we further reveal that there is no evident evolution of the band structure in both compounds and the experimental spectra can be well reproduced by the nonmagnetic calculations, together suggesting a negligible effect of the magnetism on their electronic structures and a possibly weak coupling between the localized 4$f$ electrons and the itinerant conduction electrons. Our results offer essential insights into the interactions between magnetism, electron correlations, and topological orders in the $R$Al$X$ ($R$ = light rare earth and $X$ = Si or Ge) family.
\end{abstract}

\maketitle

The last decade has witnessed the discovery of a rich variety of novel quantum states in topological semimetals \textcolor{black}{\cite{HMWeng2016,XGWan2011,ZJWang2012,HWWeng2015,SMHuang2015,AASoluyanov2015,
GChang2018,ZKLiu2014,SBorisenko2014,BQLv2015X,SXu2015S,LXYang2015,SXu2015NP,ZZhu2016,BQLv2017,QXu2015,RLou2016,LMSchoop2016,XZhang2017,XFeng2018,
RLou2018X,RLou2018npj,SXu2015arc,GBian2016NC,BQLv2015NP,NiTe2018,NiTe2019,NiTe2022,app1,app2,app3},} such as those exhibiting Dirac-fermion excitations near the points of linear band crossings close to the Fermi level ($E_F$) \textcolor{black}{\cite{ZKLiu2014,SBorisenko2014,SXu2015arc,NiTe2018,NiTe2019,NiTe2022}.} The breaking of either spatial inversion symmetry or time-reversal symmetry splits the degeneracy of the Dirac point, leading to a pair of topologically protected Weyl points \cite{XGWan2011,HWWeng2015,SMHuang2015}. The nonmagnetic (NM) Weyl fermions have been observed and intensively studied in TaAs family \cite{BQLv2015X,SXu2015S,LXYang2015,SXu2015NP,BQLv2015NP,XCHuang2015PRX},
WTe$_2$ \cite{AliMN2014,YWu2016WT}, MoTe$_2$ \cite{TamaiA2016PRX,JiangJ2017NC,AMZhang2019PRB}, etc., while the realization of magnetic Weyl semimetal state is still rare \cite{Borisenko2019YbMnBi,QWang2018CSS,DFLiu2019CSS}. Compared with the NM ones, the magnetic Weyl semimetals offer a fertile playground for the interplay between magnetism, electron correlations, and nontrivial topology, which could give rise to diverse exotic quantum phenomena, like quantum anomalous Hall effect, Majorana fermions, pair density wave, and topological axion states \cite{BAB2022review}.

It has recently been proposed that the $R$Al$X$ ($R$ = light rare earth and $X$ = Si or Ge) family with the non-centrosymmetric LaPtSi-type structure would be an ideal candidate of magnetic Weyl semimetal, where various magnetic ground states and Weyl semimetal states have been suggested by varying the rare-earth ions \cite{SYXu2017La,DSSanchez2020Pr,LyuM2020Pr,Zhao2021Pr,Zhao2021Sm,PuphalP2020Ce,HYYang2021Ce,GaudetJ2021Nd,XHYao2022Sm,
HuZX2020La,YangHY2020Pr,MengB2019PrAlGe,NgT2021La,ChangG2018cal,WangJ2022Nd}. Due to the formation of Weyl fermions by the inversion symmetry breaking prior to the magnetic transitions, the Weyl nodes are predicted to be robust and less dependent on the details of the magnetism, which acts as a simple Zeeman coupling that shifts the Weyl nodes in momentum space \cite{ChangG2018cal}. Many intriguing properties have been observed among the $R$Al$X$ family of compounds, including the topological Hall effect in CeAlGe \cite{PuphalP2020Ce}, novel anisotropic anomalous Hall effect in CeAlSi \cite{HYYang2021Ce}, large anomalous Hall conductivity and field-induced Lifshitz transition in PrAlSi \cite{LyuM2020Pr,Zhao2021Pr}, Weyl fermions driven collective magnetism in NdAlSi \cite{GaudetJ2021Nd}, and topological spiral magnetic order and non-saturated magnetoresistance in SmAlSi \cite{XHYao2022Sm,Zhao2021Sm}. Despite these substantial findings, the experiments to directly disclose the effect of $f$-electrons-induced magnetism and correlations on the nontrivial band topology are still lacking.

In order to explore the underlying Weyl physics and gain insights into its intricate interaction with the spin configuration of 4$f$ states, we present here
combined angle-resolved photoemission spectroscopy (ARPES) measurements and first-principles calculations of PrAlSi and SmAlSi crystals. We observe that their low-energy electronic structures are very similar and less sensitive to the respective magnetic transition. The band structure calculations in the paramagnetic (PM) phase can well describe the ARPES spectra. Our observations indicate a possibly negligible coupling between the well-localized 4$f$ states and the itinerant electronic states in PrAlSi and SmAlSi.

\begin{figure*}[t]
  \setlength{\abovecaptionskip}{-0.45cm}
  \setlength{\belowcaptionskip}{-0.05cm}
  \begin{center}
  \includegraphics[trim = 0mm 0mm 0mm 0mm, clip=true, width=1.9\columnwidth]{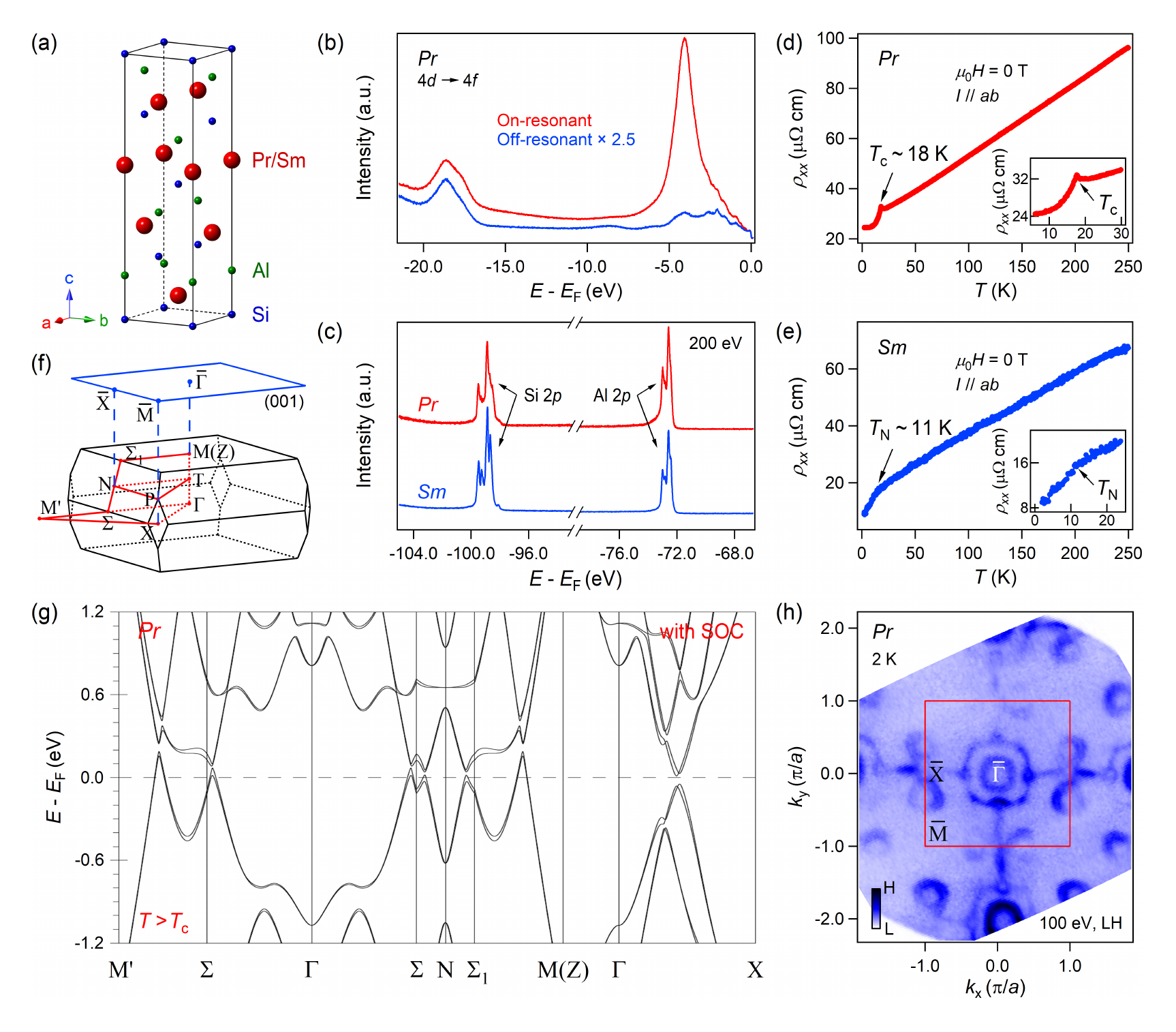}
  \end{center}
  \caption{Single crystals of PrAlSi and SmAlSi.
  (a) Schematic crystal structures of PrAlSi and SmAlSi.
  (b) Angle-integrated photoemission spectra of PrAlSi with Pr $N$ edge on-resonant (124 eV) and off-resonant (120 eV) photons, respectively.
  (c) Core-level photoemission spectra of PrAlSi and SmAlSi recorded at $h$$\nu$ = 200 eV, respectively.
  (d),(e) Temperature dependence of the resistivity $\rho_{xx}$ of PrAlSi and SmAlSi at $\mu_0$$H$ = 0 T, respectively.
  Insets show the temperature ranges close to $T_c$ and $T_N$, respectively.
  (f) Sketches of 3D BZ and (001)-surface BZ for the noncentrosymmetric $I$4$_1$/$md$ space group structure.
  (g) Calculated bulk band structure along the high-symmetry lines in the PM phase of PrAlSi including the SOC effect.
  (h) Constant-energy ARPES image of PrAlSi ($h\nu$ = 100 eV, linear horizontal polarization, $T$ = 2 K) obtained by integrating the photoemission intensity within $E_F$ $\pm$ 40 meV. The red solid curve represents the (001)-projected BZ. $a$ (= 4.22 \AA) is the in-plane lattice constant of PrAlSi.
  }
\end{figure*}

High-quality single crystals of PrAlSi and SmAlSi were synthesized using the flux method \cite{Zhao2021Pr,Zhao2021Sm}. ARPES measurements were performed using the $1^3$-ARPES end station of UE-112-PGM2 beamline at Helmholtz Zentrum Berlin BESSY-II light source. The energy and angular resolutions were set to better than 5 meV and 0.1$^{\circ}$, respectively. Samples were cleaved $\emph{in situ}$, yielding flat mirrorlike (001) surfaces. During the experiments, the sample temperature was kept at 1.5 K if not specified otherwise, and the vacuum conditions were maintained better than 8 $\times$ 10$^{-11}$ Torr.
Density functional based calculations were performed for the experimental crystal structure of PrAlSi \cite{LyuM2020Pr} using the PY linear muffin-tin orbital (LMTO) computer code \cite{book:AHY04}. We used the Perdew-Burke-Ernzerhof revised for solid parameterization of the exchange correlation potential \cite{PRCV+08}. Spin-orbit coupling (SOC) was included in the LMTO Hamiltonian at the variational step. Pr $4f^{2}$ electrons were treated as semi-core states.

\begin{figure}[t]
  \setlength{\abovecaptionskip}{-0.35cm}
  \setlength{\belowcaptionskip}{-0.1cm}
  \begin{center}
  \includegraphics[trim = 4.0mm 0mm 0mm 0mm, clip=true, width=1.04\columnwidth]{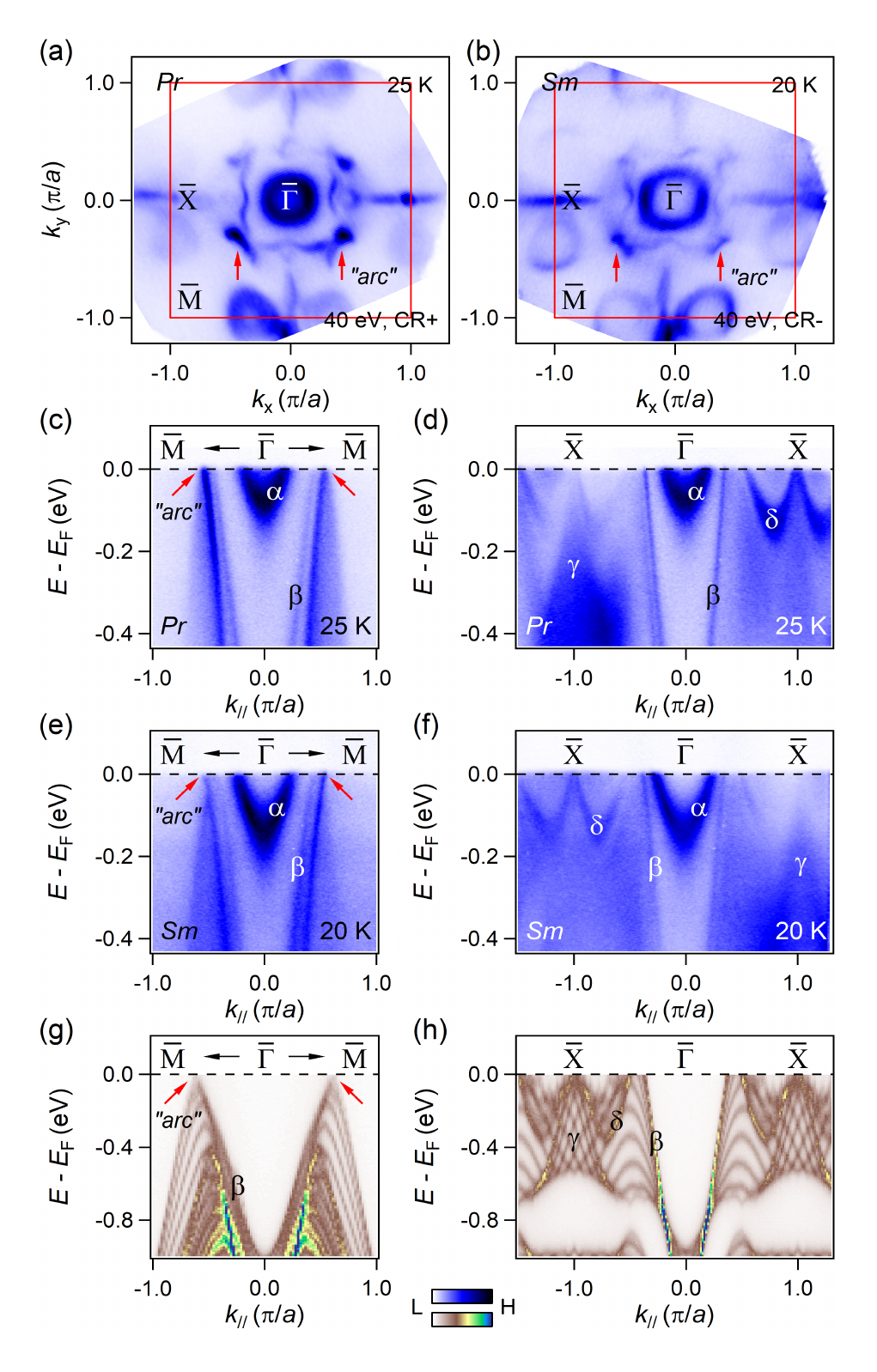}
  \end{center}
  \caption{Electronic structures of PrAlSi and SmAlSi in the PM states.
  (a),(b) ARPES intensity plots at $E_F$ of PrAlSi ($h\nu$ = 40 eV, CR$^{+}$ polarization, $T$ = 25 K) and SmAlSi ($h\nu$ = 40 eV, CR$^{-}$ polarization, $T$ = 20 K), respectively.
  The red solid curves are the (001)-surface BZs. The in-plane lattice constant of SmAlSi (= 4.16 \AA) is also denoted as $a$.
  (c),(d) Intensity plots of PrAlSi ($h\nu$ = 40 eV, CR$^{+}$ polarization) recorded along the $\bar{\Gamma}$-$\bar{M}$ and $\bar{\Gamma}$-$\bar{X}$ directions, respectively.
  (e),(f) Same as (c),(d) of SmAlSi ($h\nu$ = 40 eV, CR$^{-}$ polarization).
  (g),(h) Bulk band calculations integrated over $k_z$ from 0 to $\pi$ in the PM phase of PrAlSi taken along the $\bar{\Gamma}$-$\bar{M}$ and
  $\bar{\Gamma}$-$\bar{X}$ directions, respectively.
  }
\end{figure}

\begin{figure*}[t]
  \setlength{\abovecaptionskip}{-0.15cm}
  \setlength{\belowcaptionskip}{-0.0cm}
  \begin{center}
  \includegraphics[trim = 0mm 0mm 0mm 0mm, clip=true, width=1.92\columnwidth]{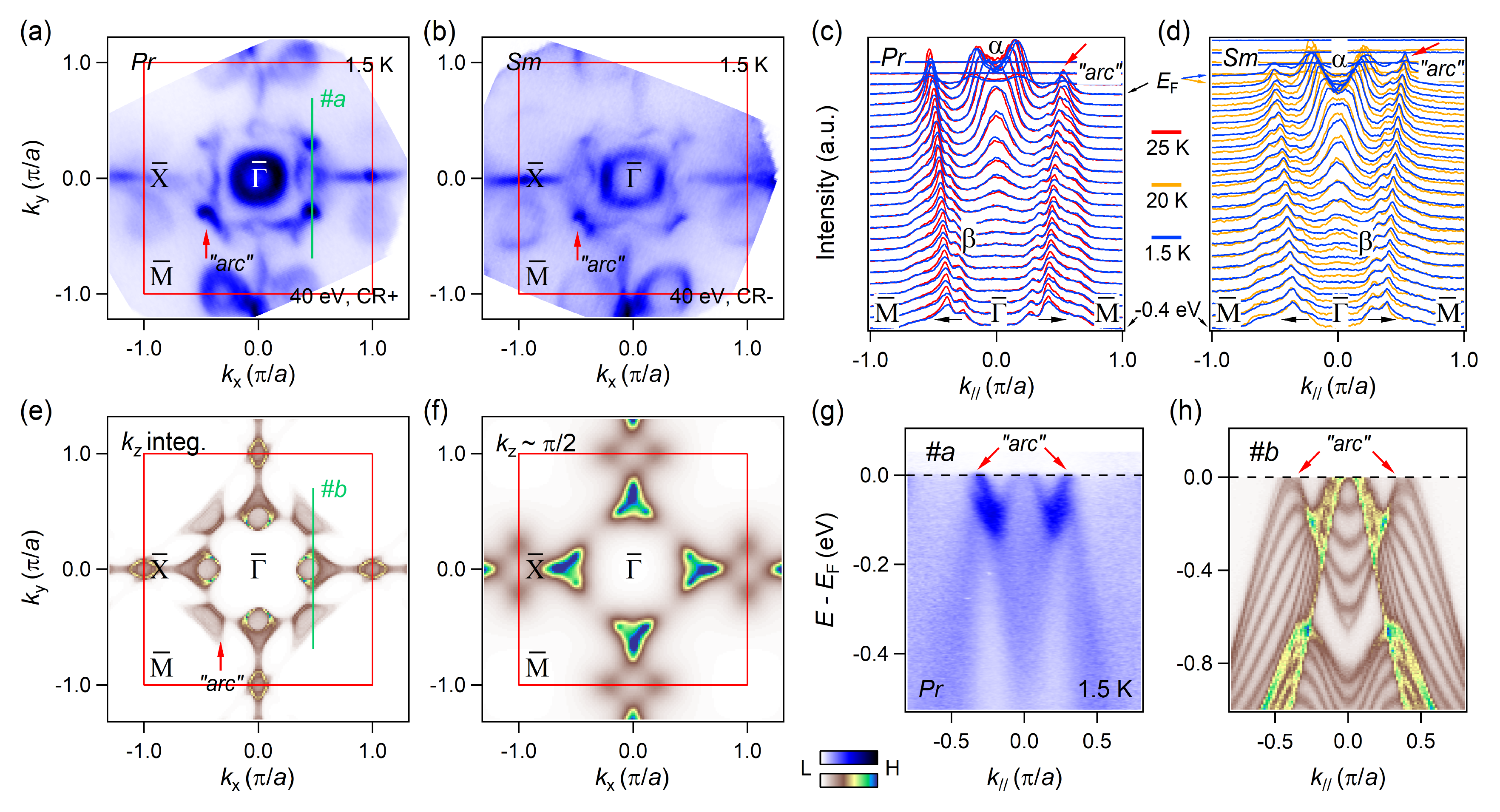}
  \end{center}
  \caption{Electronic structures of PrAlSi and SmAlSi in the magnetic states.
  (a),(b) Constant-energy maps at $E_F$ of PrAlSi ($h\nu$ = 40 eV, CR$^{+}$ polarization, $T$ = 1.5 K) and SmAlSi ($h\nu$ = 40 eV, CR$^{-}$ polarization,
  $T$ = 1.5 K), respectively.
  (c) Comparison of the MDCs along the $\bar{\Gamma}$-$\bar{M}$ direction above and below $T_c$ of PrAlSi.
  (d) Same as (c) of SmAlSi by going across $T_N$.
  (e) $K_z$-integrated FS mapping from 0 to $\pi$ in the PM state of PrAlSi by bulk calculations.
  (f) Same as (e) for $k_z$ $\sim$ $\pi$/2 [$T$-$N$-$P$ plane in Fig. 1(f)] without $k_z$ integration.
  The red solid curves in (a),(b) and (e),(f) show the (001)-projected BZs. Cuts \#a and \#b in (a) and (e) indicate the locations of the band dispersions
  in (g) (also \textcolor{black}{Fig. S2} \cite{SM}) and (h), respectively.
  (g) Intensity plot of PrAlSi ($h\nu$ = 40 eV, CR$^{+}$ polarization) measured along cut \#a in (a).
  (h) Calculated bulk band structure of the NM PrAlSi integrated over $k_z$ from 0 to $\pi$ along cut \#b in (e).
  }
\end{figure*}

As shown in Fig. 1(a), the $R$Al$X$ family crystallizes in a tetragonal structure with the non-centrosymmetric space group $I$4$_1$/$md$ (No. 109), where each stacking layer along the $c$ axis consists of one type of element. Figure 1(b) displays the resonant ARPES measurements of PrAlSi at $N$ edge of Pr element. At the Pr 4$d$ $\rightarrow$ 4$f$ resonant photon energy of 124 eV, the resonant enhancement of Pr 4$f$ ($E_F$ to $\sim$--6 eV) and 5$p$ ($\sim$--17 eV to $\sim$--20 eV) states are clearly observed compared to the off-resonant spectra with 120-eV photons. The Al 2$p$ and Si 2$p$ core-level states of PrAlSi and SmAlSi are
presented in Fig. 1(c). Their line shapes are composed of both main peaks and shoulders, indicating the existence of several sites and components of Al and Si atoms. As illustrated in Figs. 1(d) and 1(e), the underlying ferromagnetic (FM) and antiferromagnetic (AFM) transitions of PrAlSi and SmAlSi at $T_c$ $\sim$ 18 K and $T_N$ $\sim$ 11 K, respectively, are validated by the zero-field resistivity measurements, consistent with previous studies \cite{Cao2022CPL,LyuM2020Pr}. The onset of magnetic orderings below $T_c$ and $T_N$ leads to the loss of spin-disorder scattering and the decrease of the electrical resistivity, which manifests as an anomaly in the $\rho_{xx}$-$T$ curves.

Figure 1(f) presents the bulk and (001)-projected surface Brillouin zones (BZs) of the $R$Al$X$ family. Along the high-symmetry lines marked out in Fig. 1(f)
(red curves), we plot the overall bulk band structure calculations in the PM state of PrAlSi with the SOC effect in Fig. 1(g).
The semimetallic ground state is demonstrated by the crossing of conduction and valance bands along the $\Gamma$-$\Sigma$-$N$-$\Sigma_1$ path.
As shown in Fig. 1(h), the experimental constant-energy map at $E_F$ of PrAlSi recorded by 100-eV photons, which covers also some parts of the second BZ, can confirm the tetragonal crystalline symmetry with correct in-plane lattice parameter. The measured Fermi surface (FS) consists of the two square-like pockets around $\bar{\Gamma}$, the dumbbell-like pockets around $\bar{X}$, and the ripple-shaped FS contours across the BZ boundaries.

In order to uncover the detailed electronic properties of the Weyl semimetal phases in PrAlSi and SmAlSi, we carry out high-resolution ARPES measurements in their PM states at first, where the Weyl physics is determined by the inversion symmetry breaking as suggested by earlier studies \cite{ChangG2018cal,SYXu2017La,Madhab2022Ce}.
As shown in Figs. 2(a) and 2(b), apart from the small differences in the size of certain pockets, the analogous FS topologies are shared between PrAlSi and SmAlSi in the first BZs \textcolor{black}{(see Fig. S1 of Supplemental Material \cite{SM} for their similar FSs under the same photon polarization).} In comparison to that in Fig. 1(h), one can clearly observe an additional FS sheet (indicated by red arrows) close to the corner of the outer pocket around $\bar{\Gamma}$ in both PrAlSi and SmAlSi. This contour has also been previously reported in LaAlGe \cite{SYXu2017La}, PrAlGe \cite{DSSanchez2020Pr}, and CeAlSi \cite{Madhab2022Ce}, where it is proposed as the unclosed ``Fermi arc" connecting the topological Weyl fermions. The ``arc"-like feature is further
suggested to have distinct winding types between the FM and NM phases of PrAlGe \cite{DSSanchez2020Pr}, with the asymmetric and symmetric forms across the $\bar{\Gamma}$-$\bar{M}$ line, respectively. Nevertheless, our measurements of the NM PrAlSi [Fig. 2(a)] reveal an evident asymmetry of the ``arc", similar to that in the NM phase of CeAlSi, which also hosts the FM ground state \cite{Madhab2022Ce}. These results may raise concern about the previous assignment of the Fermi arcs in experiments and will be discussed in more detail later.

In Figs. 2(c) and 2(d) [Figs. 2(e) and 2(f)], we present the near-$E_F$ band structure in the PM phase of PrAlSi (SmAlSi) recorded along the $\bar{\Gamma}$-$\bar{M}$ and $\bar{\Gamma}$-$\bar{X}$ directions, respectively. It can be seen that the outer faint band ($\beta$) around $\bar{\Gamma}$ and the neighbouring hole band (red arrow), which defines the ``arc"-like FS, are clearly separated from each other along the $\bar{\Gamma}$-$\bar{M}$ direction [Figs. 2(c) and 2(e)]. As shown in Figs. 2(d) and 2(f), due to the effect of ARPES matrix element, the Dirac-like hole band ($\gamma$) and multiple electron bands ($\delta$) are observable at counter $\bar{X}$ points of the BZ respectively. On switching the photon polarization from \textcolor{black}{left-handed circular (CR$^{+}$)} [Fig. 2(d)] to \textcolor{black}{right-handed circular (CR$^{-}$)} [Fig. 2(f)], the observations are exchanged accordingly. Since the $k_z$ component is not strictly conserved in ARPES measurements, and thus cause the $k_z$ broadening effect, which is found to be significant in the vacuum ultraviolet (VUV) regime, the ARPES spectra reflect the electronic states integrated over a certain $k_z$ region of the bulk BZ \cite{Kumigashira1998kz,RLou2018X}. Therefore, we perform the bulk band calculations by considering the $k_z$ integration from 0 to $\pi$. The corresponding results in the PM phase of PrAlSi are shown in Figs. 2(g) and 2(h). The simulations can well reproduce most of the experimental band dispersions including the ``arc"-like feature along the $\bar{\Gamma}$-$\bar{M}$ direction, where the $\beta$ band and the ``arc"-related band could be degenerate near $E_F$ in the calculations [Fig. 2(g)]. The good consistency between experiments and theory suggests the bulk origin of these ARPES spectra and the presence of intrinsic $k_z$ projections. The intense electron band ($\alpha$) at $\bar{\Gamma}$ is not captured in the bulk calculations and thus could be the surface state,
\textcolor{black}{consistent with the previous observation and assignment in LaAlGe \cite{SYXu2017La}, PrAlGe \cite{DSSanchez2020Pr}, and CeAlSi \cite{Madhab2022Ce}.}

To investigate the influence of FM and AFM orderings on the electronic structures of PrAlSi and SmAlSi, we now conduct ARPES measurements deep within their
magnetic states respectively, as shown in Fig. 3. Figures 3(a) and 3(b) illustrate the corresponding FS mappings at $T$ = 1.5 K with the same experimental
setups as in Figs. 2(a) and 2(b), respectively. No noticeable change is observed across $T_c$ and $T_N$, especially the ``arc"-like FSs, where the likely asymmetry of PrAlSi and symmetry of SmAlSi across the $\bar{\Gamma}$-$\bar{M}$ line still persist. We further focus on the temperature evolution of the ``arc"-like feature
by plotting the momentum distribution curves (MDCs) along the $\bar{\Gamma}$-$\bar{M}$ direction in Figs. 3(c) and 3(d). It can be clearly seen that
the MDC peak positions above and below $T_c$ and $T_N$ coincide with each other in the studied energy range, showing the temperature independence of
the ARPES spectra. These observations unambiguously demonstrate a negligible effect of the long-range magnetic orders on the conduction electrons in PrAlSi and SmAlSi.

Now we discuss the possible origin of the ``unclosed" FS sheet which is early assigned as the topological Fermi arc \cite{DSSanchez2020Pr}. Based on the present results that this ``arc"-like structure can be well described by the $k_z$-integrated bulk band calculations, we further simulate the bulk FS topologies of the (001) surface in the full 3D BZ of the PM PrAlSi. As the good agreement illustrated in Fig. 2, the theoretical FS integrated over $k_z$ from 0 to $\pi$ in Fig. 3(e) reproduces the outer square-like pocket around $\bar{\Gamma}$, the ripple-shaped FS across the BZ boundary, and particularly the ``arc"-like FS contour.
The dumbbell-like pockets around $\bar{X}$ can be seen in the calculated FS for $k_z$ = $\pi$/2 [$T$-$N$-$P$ plane in Fig. 1(f)] without $k_z$ integration in
Fig. 3(f). The surface state nature of the inner pocket at $\bar{\Gamma}$ can be further confirmed by its absence in Figs. 3(e) and 3(f). To solely study the ``arc"-related hole band without any possible interference from the $\beta$ band, we record the ARPES spectra along cut \#a [green line in Fig. 3(a)], as shown in Fig. 3(g). By comparing with the bulk calculations fully integrated over $k_z$ taken along cut \#b in Fig. 3(e) [Fig. 3(h)], one can obtain a high consistency again. Although our photon-energy-dependent measurements find that the ``arc"-like feature shows similar $\Delta$$k_F$ $\sim$ 0.4 \AA$^{-1}$ from 30 to 60 eV (see \textcolor{black}{Fig. S2} of Supplemental Material \cite{SM}) as reported in earlier studies \cite{DSSanchez2020Pr,Madhab2022Ce}, due to the concern that the observed ``arcs" may be actually the bulk states, it is inferred that the ARPES intensity suffers from the large $k_z$-broadening effect. This is judged from not only the good agreement between experiments and bulk calculations, but also the observation of similar band structure in a wide VUV-photon energy range, as seen from Figs. 1(h) ($h\nu$ = 100 eV), 3(a)-3(b) ($h\nu$ = 40 eV), and \textcolor{black}{S3} ($h\nu$ = 50 eV) \cite{SM}.
We suspect that the theoretically predicted topological Fermi arc is likely hidden by the ``arc"-like feature revealed here. Future investigations to differentiate between the surface and bulk contributions will be required.

Last but not least, we now turn to the potential implications of the temperature-independent electronic structures in PrAlSi and SmAlSi. It has been previously
proposed that the occupied $f$-electron states in the $R$Al$X$ family (excluding the La-based compounds) are spin polarized and the magnetic interactions between the local moments of $f$ electrons can lead to the long-range magnetic ordering, which, in turn, serves as an effective Zeeman field to make the near-$E_F$ conduction bands also spin polarized \cite{ChangG2018cal,Zhao2021Sm}. Whereas, our results uncover a minor impact of the magnetic orders on the low-energy electronic states of PrAlSi and SmAlSi. This decoupling relationship implies that the 4$f$ electrons in these two compounds could be well localized and their interactions with the itinerant conduction electrons are likely negligible or in the weak-coupling regime.
\textcolor{black}{This weak hybridization phenomenon could be further revealed by the resonant ARPES measurements. As shown in \textcolor{black}{Fig. S4} \cite{SM}, no resonant enhancement is observed for the conduction states near $E_F$, the overall intensities of photoemission signals measured under on-resonant and off-resonant photons are comparable to each other.}
\textcolor{black}{In a recent ARPES study of the AFM Kondo lattice CeRh$_2$Si$_2$, the surface- and bulk-type Ce-4$f$ electrons have been demonstrated to show distinct hybridization phenomena with the itinerant electrons, representing the weakly and strongly hybridized 4$f$ states, respectively \cite{VyalikhDV2016CeRh2Si2}. Possibly similar to CeRh$_2$Si$_2$, the well-localized 4$f$ states uncovered here may originate from the surface-type Pr/Sm ions,
where the effect of bulk magnetism is negligible in the surface-sensitive VUV-ARPES spectra. Further bulk-sensitive photoemission studies will be required.}

On the other side, it has been recently reported that the magnetic properties of these two compounds could be more complicated \cite{LyuM2020Pr,XHYao2022Sm}.
In PrAlSi, the reentrant spin glassy phases are suggested to emerge just below $T_c$ \cite{LyuM2020Pr}; in SmAlSi, the Weyl-mediated Ruderman-Kittel-Kasuya-Yosida  interactions are proposed to induce a spiral magnetic order in the ground state \cite{XHYao2022Sm}. These random and/or competing exchange interactions may cause the instability of the long-range magnetic orders, leading to another challenge of observing the magnetic impact on the electronic states as well.

In summary, we have systematically studied the electronic structures in the NM and magnetic phases of PrAlSi and SmAlSi. The low-energy band structure has almost no change when the respective long-range FM and AFM order develops in them and the measured spectra show good agreement with the NM band calculations. These facts can be attributed to the possibly negligible coupling between the well-localized 4$f$ electrons and the itinerant conduction electrons. Our observations shed light on the further research of the interplay between magnetism, correlations, and topology, which will facilitate realizing more topological quantum phenomena in the $R$Al$X$ family.

\begin{acknowledgments}
We thank Denis Vyalikh for helpful discussions.
This work was supported by the National Natural Science Foundation of China (Grants No. 11904144 and No. 12004123) and the Deutsche Forschungsgemeinschaft under Grant SFB 1143 (project C04). B. B. and S. B. acknowledge the support from BMBF via project UKRATOP. R. L., A. F., B. B., and S. B. acknowledge
the support from W{\"u}rzburg-Dresden Cluster of Excellence on Complexity and Topology in Quantum Matter-\emph{ct.qmat} (EXC 2147, project-id
390858490).
\end{acknowledgments}

\end{document}